\PassOptionsToPackage{unicode}{hyperref}
\PassOptionsToPackage{hyphens}{url}
\PassOptionsToPackage{dvipsnames,svgnames,x11names}{xcolor}
\documentclass[
  12pt]{article}

\usepackage{amsmath,amssymb}
\usepackage{iftex}
\ifPDFTeX
  \usepackage[T1]{fontenc}
  \usepackage[utf8]{inputenc}
  \usepackage{textcomp} 
\else 
  \usepackage{unicode-math}
  \defaultfontfeatures{Scale=MatchLowercase}
  \defaultfontfeatures[\rmfamily]{Ligatures=TeX,Scale=1}
\fi
\usepackage{lmodern}
\ifPDFTeX\else  
\fi
\IfFileExists{upquote.sty}{\usepackage{upquote}}{}
\IfFileExists{microtype.sty}{
  \usepackage[]{microtype}
  \UseMicrotypeSet[protrusion]{basicmath} 
}{}
\makeatletter
\@ifundefined{KOMAClassName}{
  \IfFileExists{parskip.sty}{%
    \usepackage{parskip}
  }{
    \setlength{\parindent}{0pt}
    \setlength{\parskip}{6pt plus 2pt minus 1pt}}
}{
  \KOMAoptions{parskip=half}}
\makeatother
\usepackage{xcolor}
\makeatletter
\ifx\paragraph\undefined\else
  \let\oldparagraph\paragraph
  \renewcommand{\paragraph}{
    \@ifstar
      \xxxParagraphStar
      \xxxParagraphNoStar
  }
  \newcommand{\xxxParagraphStar}[1]{\oldparagraph*{#1}\mbox{}}
  \newcommand{\xxxParagraphNoStar}[1]{\oldparagraph{#1}\mbox{}}
\fi
\ifx\subparagraph\undefined\else
  \let\oldsubparagraph\subparagraph
  \renewcommand{\subparagraph}{
    \@ifstar
      \xxxSubParagraphStar
      \xxxSubParagraphNoStar
  }
  \newcommand{\xxxSubParagraphStar}[1]{\oldsubparagraph*{#1}\mbox{}}
  \newcommand{\xxxSubParagraphNoStar}[1]{\oldsubparagraph{#1}\mbox{}}
\fi
\makeatother

\usepackage{longtable,booktabs,array}
\usepackage{calc} 
\usepackage{etoolbox}
\makeatletter
\patchcmd\longtable{\par}{\if@noskipsec\mbox{}\fi\par}{}{}
\makeatother
\IfFileExists{footnotehyper.sty}{\usepackage{footnotehyper}}{\usepackage{footnote}}
\makesavenoteenv{longtable}
\usepackage{graphicx}
\makeatletter
\def\maxwidth{\ifdim\Gin@nat@width>\linewidth\linewidth\else\Gin@nat@width\fi}
\def\maxheight{\ifdim\Gin@nat@height>\textheight\textheight\else\Gin@nat@height\fi}
\makeatother
\setkeys{Gin}{width=\maxwidth,height=\maxheight,keepaspectratio}
\makeatletter
\def\fps@figure{htbp}
\makeatother

\makeatletter
\@ifpackageloaded{caption}{}{\usepackage{caption}}
\AtBeginDocument{%
\ifdefined\contentsname
  \renewcommand*\contentsname{Table of contents}
\else
  \newcommand\contentsname{Table of contents}
\fi
\ifdefined\listfigurename
  \renewcommand*\listfigurename{List of Figures}
\else
  \newcommand\listfigurename{List of Figures}
\fi
\ifdefined\listtablename
  \renewcommand*\listtablename{List of Tables}
\else
  \newcommand\listtablename{List of Tables}
\fi
\ifdefined\figurename
  \renewcommand*\figurename{Figure}
\else
  \newcommand\figurename{Figure}
\fi
\ifdefined\tablename
  \renewcommand*\tablename{Table}
\else
  \newcommand\tablename{Table}
\fi
}
\@ifpackageloaded{float}{}{\usepackage{float}}
\floatstyle{ruled}
\@ifundefined{c@chapter}{\newfloat{codelisting}{h}{lop}}{\newfloat{codelisting}{h}{lop}[chapter]}
\floatname{codelisting}{Listing}

\makeatother
\makeatletter
\@ifpackageloaded{caption}{}{\usepackage{caption}}
\@ifpackageloaded{subcaption}{}{\usepackage{subcaption}}
\makeatother

\ifLuaTeX
  \usepackage{selnolig}  
\fi
\usepackage[]{natbib}
\usepackage{bookmark}
\usepackage{pdfpages}
\newif\ifarXiv
\arXivtrue

\IfFileExists{xurl.sty}{\usepackage{xurl}}{} 
\hypersetup{
  pdftitle={Title},
  pdfauthor={Author 1; Author 2},
  pdfkeywords={3 to 6 keywords, that do not appear in the title},
  colorlinks=true,
  linkcolor={blue},
  filecolor={Maroon},
  citecolor={Blue},
  urlcolor={Blue},
  pdfcreator={LaTeX via pandoc}}

\newcommand{\blue}[1]{\textcolor{blue}{#1}}
\newcommand{\red}[1]{\textcolor{red}{#1}}

\newcommand{\anon}{1}

\begin{document}

\def\spacingset#1{\renewcommand{\baselinestretch}%
{#1}\small\normalsize} \spacingset{1}


\if1\anon
{
  \title{\bf History matching for functional data: Application to tsunami warnings in the Indian Ocean}
  \author{Ryuichi Kanai\thanks{
    Corresponding Author((Email: ryuichi.kanai.16@alumni.ucl.ac.uk)}\hspace{.2cm}\\
    Department of Statistical Science, University College London,\\ London, United Kingdom\\
    Alan Turing Institute, London, United Kingdom\\\\
    Nicol\'{a}s Hern\'{a}ndez  \\
    Data Science, Statistics and Probability Centre, \\
    School of Mathematical Sciences, \\Queen Mary University of London, London, United Kingdom\\\\
    Devaraj Gopinathan  \\
    Advanced Research Computing Centre, University College London, \\London, United Kingdom \\
    and \\\\
    Serge Guillas \\
    Department of Statistical Science, University College London,\\ London, United Kingdom
    }
  \maketitle
} \fi

\if0\anon
{
  \bigskip
  \bigskip
  \bigskip
  \begin{center}
    {\LARGE\bf History matching for functional data. Application to tsunami warnings in the Indian Ocean.}
\end{center}
  \medskip
} \fi

\bigskip
\begin{abstract}
Traditional history matching (HM) is widely used as a computationally tractable alternative to Bayesian calibration for ruling out implausible regions of the input space of expensive computer models. Its standard formulation, however, is primarily tailored to scalar or finite-dimensional vector outputs, whereas observations in many physical systems are naturally functional, taking the form of time series or spatial fields. We propose Functional History Matching (FHM), an extension of HM to functional data. FHM defines implausibility through a function-level discrepancy measure together with a scalarized uncertainty term that combines observation error, model discrepancy, and emulator uncertainty. To quantify functional mismatch, we introduce a Wiener-process-based random projection criterion that is sensitive to differences in overall shape and temporal alignment, and we allow derivative information to be incorporated when additional discriminatory power is required. For functional emulation, we adopt a multi-output Gaussian process framework implemented through the Outer Product Emulator, which yields predictive means and uncertainty summaries for functional outputs at practical computational cost. We further motivate threshold selection through a conservative, distribution-free argument based on Chebyshev's inequality. In a synthetic tsunami forecasting case study, FHM progressively contracts the not-ruled-out-yet region and yields more concentrated downstream coastal predictions than landmark-based HM.
\end{abstract}

\noindent%
\vfill

\newpage
\spacingset{1.8} 
\section{Introduction}

Scientific and engineering decision-making based on expensive numerical simulators requires efficient exploration of the input space under limited computational budgets. 
History Matching (HM) addresses this problem by using an emulator-based implausibility measure to rule out regions of the input space that are clearly inconsistent with observations, while retaining the Not--Ruled--Out--Yet (NROY) region for further exploration across successive waves of screening \citep{craig1996, vernon2014galaxy}.
The definition of implausibility is therefore central to both the rate of contraction of the search space and the risk of discarding the true setting.

Most existing HM formulations are designed for scalar or finite-dimensional vector outputs, whereas observations in many scientific applications are naturally functional data, taking the form of continuous functions such as time series or spatial fields. Extending HM directly to such outputs, however, is nontrivial.
First, the emulator should preserve dependence and smoothness along the output domain, rather than treating the output merely as a high-dimensional vector. 
Second, the framework requires a measure of functional distance to evaluate discrepancies, reflecting differences in shape, phase and, when useful, derivative information.
Third, derivative-based features require care because numerical differentiation can amplify noise in discretely observed functions.
Compared to the conventional scalar-output setting, these issues introduce additional methodological and computational challenges.


In this paper, we address the above challenges by proposing the \emph{Functional History Matching} (FHM) approach, which extends HM to functional data and provides a practical procedure for progressively screening the input parameter space while treating model outputs as functions. The key components of the proposed framework are as follows: (1) We define an implausibility measure based on a functional discrepancy and aggregated uncertainty to sequentially rule out inconsistent input regions, providing an implementation template for HM with functional outputs by treating observation error and model discrepancy jointly. (2) We introduce a Wiener-process-based random-projection distance-like score \citep{dasgupta2013experiments}, as a distance-like measure that is sensitive to phase shifts and waveform shape differences, and, when necessary, combine it with derivative features so that rejection decisions can be made using both function values and derivatives.
(3) We present a distribution-free implausibility rejection threshold design based on Chebyshev's inequality and organize it as an operational guideline for practice.
(4) We adopt a multi-output Gaussian process (MOGP) framework for emulating functional outputs, implemented through the Outer Product Emulator (OPE) \citep{rougier2008efficient},  which provides a structured covariance representation and enables efficient screening over dense candidate sets. 
(5) We evaluate the performance of FHM against conventional scalar-output HM through a tsunami numerical model case study.

The remainder of this paper is organized as follows. In Section 2, we describe the FHM framework. In Section 3, we present a case study using a tsunami numerical model. Section 4 concludes with a discussion of the main findings and possible extensions.
%
%
%
%
%
%
%
%
%
%
\section{Framework of Functional History Matching}
\label{sec:FHM}

In this section, we summarize the design elements required to make HM applicable to functional outputs and formulate the overall framework of FHM.
%
%
%
%
%
%
\subsection{Review of Conventional History Matching}
\label{subsec:HM}

We briefly review conventional HM for scalar or finite-dimensional vector outputs \citep{vernon2014galaxy, vernon2018bayesian}.
An observation $z$ is represented as the sum of the true system output $y$ and an observation error $\phi$,
$
  z = y + \phi .
$
The true output $y$ is approximated by an expensive numerical simulator $f(\cdot)$.
To account for the discrepancy between the simulator and reality, we introduce a model discrepancy term $\delta$,
and describe the system using an input parameter vector
$\boldsymbol{\theta}\in\Theta\subset\mathbb{R}^p$ as
$
  y = f(\boldsymbol{\theta}) + \delta .
$
When evaluating $f(\cdot)$ is computationally costly and the input space $\Theta$ is high-dimensional,
an exhaustive search is infeasible in practice. Therefore, an emulator $f_{\mathrm{emu}}(\cdot)$
is constructed from a limited number of simulator runs, introducing an emulator uncertainty term $\eta(\boldsymbol{\theta})$ to approximate z as follows:
\begin{equation}
  z = f_{\mathrm{emu}}(\boldsymbol{\theta}) + \eta(\boldsymbol{\theta}) + \delta + \phi .
  \label{eq:z_femu_eta}
\end{equation}

The aim of HM is not to estimate an optimum or a posterior distribution, but rather to
sequentially rule out regions that are clearly inconsistent with observations and to identify
the Not--Ruled--Out--Yet (NROY) region.
In HM, the reduction is carried out iteratively in stages, hereafter referred to as waves.
At waves $w=1,2,\dots$, the search region is reduced according to
$
  \Theta_w \subset \Theta_{w-1} \subset \cdots \subset \Theta_1 \subset \Theta .
$
To achieve this reduction, HM employs an implausibility measure $I(\boldsymbol{\theta})$,
which scales the mismatch between prediction and observation by the associated uncertainty.

Given an observation vector $\boldsymbol{z}$ and an emulator output
$\boldsymbol{f}_{\mathrm{emu}}(\boldsymbol{\theta})$,
let $\mathbb{E}\{\boldsymbol{f}_{\mathrm{emu}}(\boldsymbol{\theta})\}$ denote the predictive mean, and
let $\mathrm{Var}(\boldsymbol{\phi})$, $\mathrm{Var}(\boldsymbol{\delta})$, and
$\mathrm{Var}(\boldsymbol{\eta}(\boldsymbol{\theta}))$ represent the variances associated with the
observation error, model discrepancy, and emulator uncertainty, respectively.
Then, using a suitable norm $\|\cdot\|$, the implausibility can be defined as:
\begin{equation}
  I(\boldsymbol{\theta})
  =
  \frac{
    \left\|
      \boldsymbol{z} - \mathbb{E}\{\boldsymbol{f}_{\mathrm{emu}}(\boldsymbol{\theta})\}
    \right\|
  }{
    \left\{
      \mathrm{Var}(\boldsymbol{\phi})
      + \mathrm{Var}(\boldsymbol{\delta})
      + \mathrm{Var}(\boldsymbol{\eta}(\boldsymbol{\theta}))
    \right\}^{1/2}
  } .
  \label{eq:I_vector}
\end{equation}
A threshold $T>0$ declares $\boldsymbol{\theta}$ implausible when $I(\boldsymbol{\theta})>T$, in which case it is ruled out, while points satisfying $I(\boldsymbol{\theta})\le T$ are retained as NROY and carried forward to the next wave.
Consequently, the design of $I(\boldsymbol{\theta})$ and the choice of $T$ determine both the rate of NROY contraction and the risk of incorrectly discarding the true setting.
%
%
%
\subsection{Formulation of Functional History Matching}
\label{subsec:FHM_formulation}

In many scientific and engineering applications, observations and simulation outputs are obtained as
functions such as time series waveforms or spatial fields. In such settings, it is natural to adopt the
framework of functional data analysis (FDA) \citep{wang2016functional}.
However, directly applying conventional HM to functional outputs is challenging, and it requires
emulation methods that preserve correlation structures and smoothness along the output domain, and
a principled design to measure the functional mismatch between observations and predictions.

In FHM, we denote the observed function by $z(\cdot)$, the predictive mean of a functional emulator by
$\mu(\boldsymbol{\theta},\cdot)$, and $d\{z,\mu(\boldsymbol{\theta})\}$ represents the discrepancy between them. In Section~\ref{sec:RP-distance} we introduce a distance-type score based on random projections to reflect both phase differences and shape differences.
Here, $d(\cdot,\cdot)$ is not restricted to be a strict metric; rather, it may include a distance-like score
that provides a stable measure of disagreement between functions. Moreover, since the observation error $\mathrm{Var}(\phi)$ and model discrepancy $\mathrm{Var}(\delta)$ are often
difficult to identify separately in practice, we treat only their sum as a combined scale parameter,
$
  \sigma^2_{\mathrm{obs+md}}
  =
  \mathrm{Var}(\phi)+\mathrm{Var}(\delta),
$
while the emulator uncertainty is represented by $\sigma^2_{\mathrm{emu}}(\boldsymbol{\theta})$,
which is constructed from the predictive variance.

Based on these ingredients, we define the FHM implausibility as
\begin{equation}
  I(\boldsymbol{\theta})
  =
  \frac{
    d\{z,\mu(\boldsymbol{\theta})\}
  }{
    \left\{
      \sigma^2_{\mathrm{obs+md}}
      +
      \sigma^2_{\mathrm{emu}}(\boldsymbol{\theta})
    \right\}^{1/2}
  }.
  \label{eq:I_FHM}
\end{equation}
At each wave $w$, we evaluate \eqref{eq:I_FHM} over a dense candidate set and retain the points satisfying
$I(\boldsymbol{\theta})\le T$ as the NROY region $\Theta_w$. 
Additional simulator runs are then generated from $\Theta_w$ using a space-filling design, and the emulator is updated accordingly.
In practical multi-wave HM, the outputs entering the implausibility may be adapted across waves, with outputs that are difficult to emulate over the full input space deferred until the NROY region has contracted sufficiently for adequate emulation \citep{vernon2018bayesian}. 
In the present paper, however, we keep the functional-output construction fixed across waves in order to isolate the effect of the proposed FHM formulation in a transparent case-study setting.
Repeating this procedure yields a multi-wave screening algorithm in which the NROY region is progressively contracted. In practice, the procedure is terminated when the size of the NROY region and the quantities of interest stabilise across successive waves.
%
%
%
%
%
%
\subsection{MOGP Emulators and OPE}
\label{sec:OPE}

In each wave of FHM, instead of running an expensive simulator directly over the entire candidate set
$\boldsymbol{\theta}\in\Theta$, we construct a functional emulator from simulation outputs obtained at a limited
number of training inputs, and then rapidly evaluate predictive means and uncertainties over a large number of
candidate points.
The present study focuses on functional outputs, and thus it is
desirable to employ an emulator that provides predictive distributions while preserving correlation structures
along the output domain.
%
%
%
%
%
%
A fundamental framework that meets this requirement is the Multi-Output Gaussian Process (MOGP)  \citep{LIU2018102}.  

One of the key advantages of MOGPs is that they allow one to explicitly incorporate correlation structures along the output domain through a joint covariance function. However, in a general MOGP, this leads to a massive full covariance matrix, $\Sigma$, of size $Nq \times Nq$, yielding a training computational cost that can reach $\mathcal{O}((Nq)^{3})$ \citep{LIU2018102}. Since FHM requires emulator updates across waves, such a general-form MOGP is not practical from the standpoint of training cost. Recent methods \citep{seidman2025vppe} combine the scaled Vecchia approximation method \citep{katzfuss2022scaled} with the Parallel Partial Emulation \citep{gu2016parallel} to gain significant efficiency, especially in spatio-temporal contexts.



%
%
%
%
%
\subsubsection{Structure and computational efficiency of OPE}

To overcome the computational bottleneck of a general MOGP while preserving the correlation
structure and uncertainty of functional outputs as time series, we employ the OPE \citep{rougier2008efficient}.
The OPE represents the mean structure as an outer product of basis functions defined on the input and output domains, and enables large-scale computation by assuming a separable covariance structure for
the residual process. It has already been successfully used for the functional emulation of tsunami models \citep{sarri2012statistical, guillas2018functional}.

Let $\boldsymbol{g}(\boldsymbol{\theta})\in\mathbb{R}^{v_r}$ be a regression vector on the input space and
$\boldsymbol{s}(t)\in\mathbb{R}^{v_s}$ be a regression vector on the output domain. Using a coefficient matrix
$B\in\mathbb{R}^{v_r\times v_s}$, we model
\begin{equation}
  y(t;\boldsymbol{\theta})
  =
  \boldsymbol{g}(\boldsymbol{\theta})^\top B\,\boldsymbol{s}(t)
  + \varepsilon(t;\boldsymbol{\theta}),
  \label{eq:ope_mean}
\end{equation}
where $\varepsilon(t;\boldsymbol{\theta})$ denotes a residual process around the mean. The key assumption of the OPE is to factorize the residual covariance into the product of an input-domain kernel $K_{\theta}$ and an output-domain kernel $K_t$:
\begin{equation}
  \mathrm{Cov}\!\bigl(\varepsilon(t;\boldsymbol{\theta}),\varepsilon(t';\boldsymbol{\theta}')\bigr)
  =
  k_{\theta}(\boldsymbol{\theta},\boldsymbol{\theta}')\,k_{t}(t,t').
  \label{eq:ope_sep}
\end{equation}
Under this separability assumption, the full joint covariance matrix can be handled via a Kronecker structure, $\Sigma \approx K_{\theta} \otimes K_{t}$, which substantially reduces the computational burden for matrix inversion and determinant evaluation.
As a result, computations that scale as $\mathcal{O}((Nq)^3)$ for a general MOGP can be reduced to roughly
$\mathcal{O}(N^3+q^3)$.
This reduction makes it feasible to perform iterative screening over a large candidate set using realistic
computational resources.
%
%
%
%
%
%
\subsubsection{Predictive distribution and the use of uncertainty in FHM}

Prediction at a new input $\boldsymbol{\theta}$ under the OPE is provided as a probability distribution for the output vector on the discrete grid, $\boldsymbol{y}(\boldsymbol{\theta})\in\mathbb{R}^q$ \citep{rougier2008efficient}.
By marginalizing over the coefficient matrix $B$ and the residual variance, the prediction follows a multivariate Student-$t$ predictive distribution of the form
\begin{equation}
  \boldsymbol{y}(\boldsymbol{\theta})
  \mid \mathcal{D}
  \sim
  t_{\nu}\!\bigl(\boldsymbol{\mu}(\boldsymbol{\theta}),\,\Sigma(\boldsymbol{\theta})\bigr),
  \label{eq:ope_pred_t}
\end{equation}
where $\boldsymbol{\mu}(\boldsymbol{\theta})$ is the predictive mean, $\Sigma(\boldsymbol{\theta})$ is the
predictive covariance including correlations across output locations, and $\nu$ is the degrees of freedom.  

In the FHM implausibility \eqref{eq:I_FHM}, the denominator requires an emulator uncertainty term
$\sigma^2_{\mathrm{emu}}(\boldsymbol{\theta})$.
Since the OPE provides the full output covariance $\Sigma(\boldsymbol{\theta})$, pointwise predictive variances over the output domain can be obtained from its marginal variances (diagonal elements).
To balance approximation accuracy and computational cost for large candidate sets, we summarize these pointwise variances into a representative scalar scale by averaging them over the output domain:
\begin{equation}
  \sigma^2_{\mathrm{emu}}(\boldsymbol{\theta})
  =
  \frac{1}{q}\sum_{j=1}^q \Sigma_{jj}(\boldsymbol{\theta}).
  \label{eq:sigma_emu_mean_diag}
\end{equation}
This quantity is used as the emulator uncertainty term in the denominator of the FHM implausibility.

When derivative series are required, derivative curves can be generated by sampling from the OPE predictive distribution \eqref{eq:ope_pred_t} and differentiating them, as detailed in Section \ref{subsec:derivative_extension}.
The uncertainty scale for derivatives is then evaluated from the variability of these derivative samples.
In this way, the OPE consistently provides both the predictive mean of functional outputs and an uncertainty scale, enabling high-density screening within FHM at a practical computational cost.
%
%
%

\subsection{Random Projections of Functional Distances and Derivative Features}
\label{sec:RP-distance}

In this subsection, we specify the numerator of the implausibility measure
in \eqref{eq:I_FHM}, namely
$d\{z,\mu(\boldsymbol{\theta})\}$,
as a distance-like score constructed via random projections of the functional outputs and their derivatives.
Let $z(t)$ denote the observed curve (or its smoothed version $\tilde z(t)$), and let $\mu(\boldsymbol{\theta},t)$ be the predictive mean of the functional emulator, with $t \in \mathcal{T}$.
We consider the discrepancy function
$
  f_{\boldsymbol{\theta}}(t)
  =
  \tilde z(t) - \mu(\boldsymbol{\theta},t).
$
In FHM, the role of the numerator is to summarize the magnitude of this discrepancy function
$f_{\boldsymbol{\theta}}$ into a scalar quantity,
so that candidate points $\boldsymbol{\theta}$ can be ranked for rejection.
%
%
%
\subsubsection{Distances for Functional Data and the Role of Random Projection}
A common choice for measuring similarity in function spaces is the $L^{p}$ norm.
It has been pointed out, however, that these norms may fail to adequately capture phase variation and the structure of regions where differences occur \citep{marron2015functional}.
Moreover, when functions are compared on a dense grid, the output dimension becomes large, which is disadvantageous both in terms of norm concentration and computational cost.
Hence, in practice, some form of dimension reduction is almost indispensable.

A representative dimension-reduction approach is functional principal component analysis (FPCA) \citep{ramsay2005fitting}.
While FPCA provides a low-dimensional subspace that explains a large portion of the data variability, it requires computationally expensive estimation of eigenfunctions, and it does not necessarily preserve the original pairwise distances well after projection \citep{biau2008performance}.

In this study, we instead employ random projection for functional data.
Random projection is a computationally inexpensive dimension-reduction technique \citep{vempala2005random, cuesta2007random}, whose theoretical foundation relies on the Johnson--Lindenstrauss lemma \citep{william1984extensions}.
This lemma states that any set of $n$ points in a Euclidean space can be randomly projected into a space of dimension $O(\log n / \varepsilon^2)$ while approximately preserving pairwise distances within a relative error $\varepsilon \in (0,1)$ with high probability.

Furthermore, even in infinite-dimensional (separable) Hilbert spaces, Cram\'er--Wold type results imply that a family of one-dimensional projections onto random Gaussian directions can characterize the original distribution
\citep{cuesta2007sharp, cuesta2008random}.
This viewpoint is consistent with studies that use random projections to compute
depth measures for functional data \citep{cuesta2008random, nieto2016topologically}, and it aligns with the objective of FHM in providing an ordering of functional outputs.
Although a single projection is theoretically sufficient for extracting information, in practice it is recommended to average over multiple projections in order to reduce Monte Carlo variability \citep{cuesta2006random, cuesta2008random}.
%
%
%
\subsubsection{A Random-Projection Distance-Like Score Based on the Wiener Process}

Based on the discussion above, we construct random projections in a function space
$\mathcal{H}$ as follows.
Let $\{h^{(m)}(t)\}_{m=1}^M$ be a collection of random test functions on $\mathcal{T}$
(i.e., projection directions).
For each $m=1,\dots,M$, we define the linear functional
\begin{equation}
  S^{(m)}(\boldsymbol{\theta})
  =
  \int_{\mathcal{T}} f_{\boldsymbol{\theta}}(t)\, h^{(m)}(t)\,\mathrm{d}t.
  \label{eq:rp-score-f}
\end{equation}
In this study, we employ the mean absolute projection
\begin{equation}
  D_{\mathrm{val}}(\boldsymbol{\theta})
  =
  \frac{1}{M}
  \sum_{m=1}^M
    \bigl|S^{(m)}(\boldsymbol{\theta})\bigr|
  \label{eq:rp-distance-f}
\end{equation}
as a distance-like score based on function values, and use it as the numerator of the implausibility measure.

The score $D_{\mathrm{val}}(\boldsymbol{\theta})$ is nonnegative and invariant to a
sign flip of $f_{\boldsymbol{\theta}}$.
Because only finitely many projection directions are used, however, a nonzero discrepancy function can be orthogonal to all selected directions.
Hence, 
$D_{\mathrm{val}}(\boldsymbol{\theta})=0$ does not necessarily imply
$f_{\boldsymbol{\theta}}(t)\equiv 0$.
Consequently, the score does not strictly satisfy the identity-of-indiscernibles property of a metric. 
Accordingly, we interpret $d\{z,\mu(\boldsymbol{\theta})\}$ as a distance-like score that stably quantifies functional mismatch.
Since the goal of HM is not an exact identity test but a ranking of candidate inputs for rejection, such a score is sufficiently effective for discrimination.

The discrepancy captured by \eqref{eq:rp-distance-f} depends on the distributional choice of the test functions $h^{(m)}$.
For example, if $h^{(m)}$ is generated from independent Gaussian noise on the discrete grid, the induced score is directly linked to the discrete $L^2$ norm of the discrepancy vector and therefore mainly reflects overall amplitude differences.
In contrast, to capture discrepancies that are sensitive to correlation structure and cumulative behavior along the output domain, we use Brownian random projection functions.

On a discrete output grid $t_1<\dots<t_q$, the Brownian projection directions
are generated as Gaussian vectors
$
  \boldsymbol{h}^{(m)}
  =
  \bigl(h^{(m)}(t_1),\dots,h^{(m)}(t_q)\bigr)^\top
  \sim N(\boldsymbol{0}, \Sigma_{\mathrm{B}}),
  \quad
  \Sigma_{\mathrm{B}}(i,j)=\min(t_i,t_j).
$
Here, $t_i$ denotes the grid coordinate used to generate the Brownian path.
In the numerical implementation, the Brownian paths are generated on the retained simulator grid, so that the increment variance between adjacent retained time points equals the subsampling interval. The corresponding discrete inner product is evaluated on this equally spaced retained grid using unit weights; equivalently, the common grid-spacing factor is absorbed into the definition of the score. The normalized horizontal axis shown in the figures is rescaled to \([0,1]\) only for visualization.
Under this convention,
$
  S^{(m)}(\boldsymbol{\theta})
  \approx
  \boldsymbol{f}_{\boldsymbol{\theta}}^\top \boldsymbol{h}^{(m)},
  \quad
  \boldsymbol{f}_{\boldsymbol{\theta}}
  =
  \bigl(f_{\boldsymbol{\theta}}(t_1),\dots,f_{\boldsymbol{\theta}}(t_q)\bigr)^\top,
$
and we obtain
\begin{equation}
  \mathbb{E}\bigl[(S^{(m)}(\boldsymbol{\theta}))^2
    \mid \boldsymbol{f}_{\boldsymbol{\theta}}\bigr]
  =
  \boldsymbol{f}_{\boldsymbol{\theta}}^\top
  \Sigma_{\mathrm{B}}
  \boldsymbol{f}_{\boldsymbol{\theta}}.
  \label{eq:wiener_uncertainti_discrete}
\end{equation}
For the continuous formulation in \eqref{eq:rp-score-f}, the corresponding quadratic form is defined through the Brownian kernel $K_{\mathrm{B}}(t,s)=\min(t,s)$, where \(t\) and \(s\) denote the coordinates used for the Brownian projection:
\begin{equation}
  Q(f_{\boldsymbol{\theta}})
  =
  \iint_{\mathcal{T}\times\mathcal{T}}
    f_{\boldsymbol{\theta}}(t)\, K_{\mathrm{B}}(t,s)\, f_{\boldsymbol{\theta}}(s)\,
    \mathrm{d}t\,\mathrm{d}s,
  \label{eq:wiener_uncertainti_contiue}
\end{equation}
so that the continuous random projections are stochastically linked to $Q(f_{\boldsymbol{\theta}})$.
This quadratic form tends to emphasize discrepancies that accumulate along the domain (e.g., rise and decay behavior or peak timing) rather than purely local high-frequency differences.

Under the continuous formulation, $S^{(m)}(\boldsymbol{\theta})\mid f_{\boldsymbol{\theta}}$ follows a zero-mean Gaussian
distribution with conditional variance
$
  \mathbb{E}\!\left[(S^{(m)}(\boldsymbol{\theta}))^2\mid f_{\boldsymbol{\theta}}\right]
  =
  Q(f_{\boldsymbol{\theta}}).
$
Consequently, the conditional mean absolute value used in the distance-like score,
$
  \mathbb{E}\!\left[|S^{(m)}(\boldsymbol{\theta})|\mid f_{\boldsymbol{\theta}}\right],
$
satisfies
$
  \mathbb{E}\!\left[|S^{(m)}(\boldsymbol{\theta})|\mid f_{\boldsymbol{\theta}}\right]
  =
  \sqrt{\frac{2}{\pi}}\,Q(f_{\boldsymbol{\theta}})^{1/2},
$
and can therefore be interpreted as a constant multiple of $Q(f_{\boldsymbol{\theta}})^{1/2}$.
%
%
%
\subsection{Extension to Derivative Features}
\label{subsec:derivative_extension}

Rather than training a separate emulator specifically for derivative series, we obtain derivatives by drawing discrete curve samples from the OPE predictive distribution, representing them using a B-spline basis, and then differentiating the resulting functions.

The representative derivative curve (predictive mean) is obtained by differentiating the predictive mean
$\mu(\boldsymbol{\theta},t)$ of the functional output:
$
  \mu^{(1)}(\boldsymbol{\theta},t)
  =
  \frac{\partial}{\partial t}\mu(\boldsymbol{\theta},t).
$
Applying the random projection to the derivative series yields a distance-like score based on derivatives.

Uncertainty is handled as follows. From the OPE predictive distribution at a given
input $\boldsymbol{\theta}$, we generate $N_{\mathrm{MC}}$ predictive curve samples
$\{\boldsymbol{y}_{\mathrm{pred}}^{(n)}(\boldsymbol{\theta})\}_{n=1}^{N_{\mathrm{MC}}}$. After applying the same smoothing and functionalization procedure to each sample, we compute derivative series
$\{(y_{\mathrm{pred}}^{(n)})'(t_j;\boldsymbol{\theta})\}_{j=1}^q$ by differentiation. The pointwise
uncertainty for derivatives is evaluated by the sample variance
$
  \widehat{v}_j(\boldsymbol{\theta})
  =
  \mathrm{Var}_{n=1,\dots,N_{\mathrm{MC}}}\bigl((y_{\mathrm{pred}}^{(n)})'(t_j;\boldsymbol{\theta})\bigr),
$
and its average defines the scalar uncertainty measure
$
  \sigma^2_{\mathrm{emu,der}}(\boldsymbol{\theta})
  =
  \frac{1}{q}\sum_{j=1}^q \widehat{v}_j(\boldsymbol{\theta}).
$

Here, smoothing is regarded as part of the observation operator required for the stable estimation of derivatives.
In practice, it is difficult to identify the estimation error
variance induced by smoothing separately, and adding it explicitly may lead to double counting together with observation error and model discrepancy. 
Therefore, in this study we do not introduce an additional smoothing-error term into the denominator. 
The denominator includes only two uncertainty components: the combined contribution of observation error and model discrepancy, and the emulator predictive variance.
%
%
%
\subsection{Threshold Design Based on Chebyshev's Inequality}
\label{subsec:threshold}

In conventional HM, it is common to use a threshold around $T\approx 3$, motivated by the 3-sigma rule of
\citet{pukelsheim1994three}.
In that setting, the implausibility is closely related to a standardized residual obtained by centering the observation at the emulator mean and scaling by the associated uncertainty.
By contrast, in FHM, the numerator $d\{z,\mu(\boldsymbol{\theta})\}$ is a nonlinear distance-like score constructed from functional discrepancies,
and the denominator is a scalar summary of a functional uncertainty.
As a result, the correspondence with a standardized residual is no longer exact, and the distribution of $I(\boldsymbol{\theta})$ may not be unimodal.
It is therefore preferable to adopt a threshold design that does not rely on a specific distributional form.

Accordingly, we use Chebyshev's inequality as a conservative and distribution-free guide:
\begin{equation}
  \Pr(|X-\mu|\ge k\sigma)\le \frac{1}{k^2}.
  \label{eq:chebyshev}
\end{equation}
For a residual-type quantity centered by its mean and scaled by its standard deviation, this inequality implies that deviations exceeding five standard deviations occur with probability at most $1/25 = 0.04$. 
Although the FHM implausibility is not itself such a signed standardized residual, this result provides a conservative motivation for adopting $T=5$ as an operational threshold. In this sense, $T=5$ should be interpreted not as an exact probabilistic cutoff, but as a robust rejection criterion that avoids imposing a strong distributional assumption on the functional mismatch score.

Since it is generally difficult to specify $\sigma^2_{\mathrm{obs+md}}$ accurately in advance, we do not treat it as a fixed known quantity. 
Instead, we regard $\sigma^2_{\mathrm{obs+md}}$ as a scale hyperparameter and rescale it so that the proportion of points retained in the NROY region is close
to a prescribed target level (e.g., $5\%$).
This provides a data-driven calibration of the aggregate scale without requiring an arbitrary separation of observation error from model discrepancy.
%
%
%
\section{Application to a Makran Tsunami Scenario}
\label{sec:result}

Tsunamis are powerful natural disasters causing widespread devastation, e.g. the 2004 Sumatran-Andaman tsunami led to more than 200,000 casualties \citep{lay2005great} in addition to high economic losses. 
Such a high casualty count is partly due to the low reliability of the tsunami forecasting scheme at that time, making it difficult to determine tsunami heights immediately after an earthquake event.
Conventional methods have been using earthquake sources from seismological inversion to estimate tsunami wave heights. Initial estimates based on this approach can underestimate the scale of the earthquake source, leading to tsunami size estimates that are significantly lower than the actual sea surface changes \citep{wang2021review}. 
Additionally, the extremely complex relationship between earthquake sources and initial tsunamis implies that, even if earthquake sources are accurately estimated, it is currently impossible to precisely forecast initial tsunami scale and waveforms based solely on that information \citep{Okal1988, Polet_and_Kanamori_2000}.

More accurate estimation of tsunamis is achieved by directly using observations of the actual tsunami wave.
For instance, \citet{percival2014automated} infer the tsunamigenic earthquake source (or initial sea surface displacement) directly from buoy observations. However, this framework relies on pre-computed tsunami databases that often utilize linear simplifications, such as scaling wave responses linearly across different earthquake magnitudes, an approximation that is not entirely valid in complex real-world scenarios. Another limitation is that such approaches may not adequately address the need for warnings of landslide-generated tsunamis \citep{harbitz2014submarine}.

Bypassing these physical simplifications shifts the burden toward a high-dimensional search space that is exceptionally difficult to navigate under tight emergency timelines. 

\subsection{Case-study setup and emulator adequacy}
\label{subsec:result_setup}

\begin{figure}[t!]
\begin{center}
\includegraphics[width=6in]{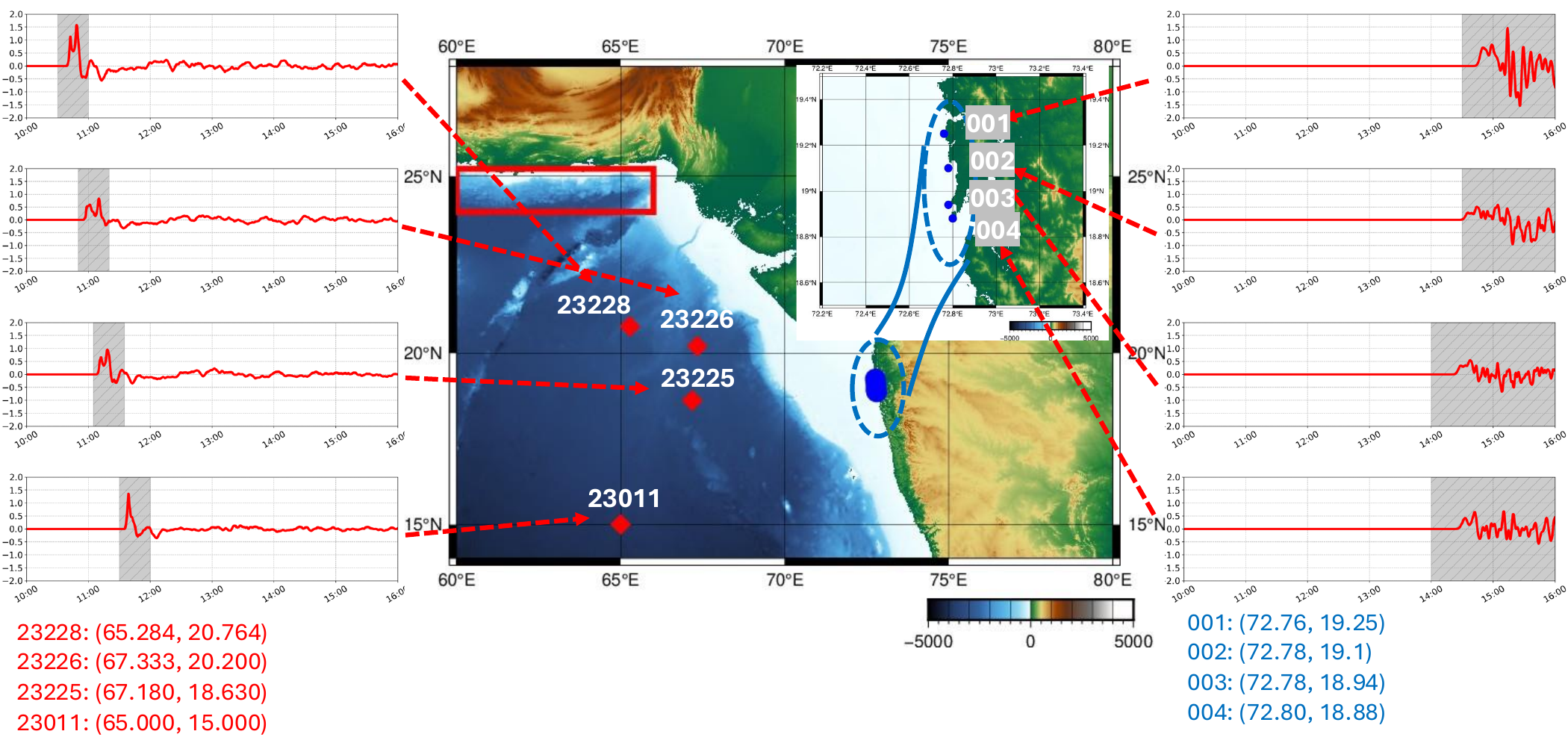}
\end{center}
\caption{\small Study region and observation locations used in the case study. DART stations correspond to red diamonds (\red{$\blacklozenge$}) and Mumbai coastal locations to blue circles (\blue{$\bullet$}). The red rectangle indicates the candidate region for the location parameters $(\theta_1,\theta_2)$ in Wave~1. 
The tsunami time series generated under the worst-case scenario (\red{\textbf{---}}) are also
shown, with tsunami height expressed in meters.
The DART series were used as pseudo-observations for FHM,
whereas the Mumbai series were used as reference series for
downstream prediction assessment.
\label{fig:case_study_overview}}
\end{figure}

For this case study, we ran the tsunami simulator JAGURS \citep{baba2015parallel} under a worst-case tsunami scenario specified for the Makran Subduction Zone (MSZ) \citep{gopinathan2021probabilistic}, and generated tsunami time series at four DART stations \citep{gonzalez1998deep} and four locations along the Mumbai coast. The tsunami time series simulated at the DART stations under this scenario were then treated as pseudo-observations, and FHM was used to sequentially reduce the input parameter space and to examine how this reduction translates into improved coastal prediction along the Mumbai coast. Figure~\ref{fig:case_study_overview} summarizes the study region, the observation locations,
and the tsunami time series generated under the worst-case scenario.
The DART time series were used as pseudo-observations for FHM,
whereas the Mumbai time series were retained as reference series
for downstream prediction assessment.
%
%
%
%
%
The input is a five-dimensional vector $\boldsymbol{\theta}$ representing the longitude and latitude of the midpoint between the initial sea-surface uplift and subsidence regions, the spatial extent of the source region in the longitudinal and latitudinal directions, and the maximum height of the initial sea-surface displacement.
In this case study, we deliberately avoided relying on strong prior restrictions based on expert knowledge, and instead specified relatively broad initial ranges for the location, shape, and amplitude parameters. This design allows us to assess whether FHM remains effective even when only limited prior information is available.

Because time-critical applications do not allow many sequential rounds of simulator runs and emulator updates, we adopted a design that combines a small number of HM waves with dense emulator-based screening. At each HM wave, the fitted emulator was used to screen $10^6$ candidate input points, and the NROY region was extracted from the resulting implausibility values. Accordingly, in this case study we adopted a two-wave FHM design. In Wave~1, 100 simulator runs were generated over the initial broad input space using a Latin hypercube design. In Wave~2, an additional 100 simulator runs were generated over a rectangular approximation to the Wave~1 NROY region.

At each observation site, the time series was restricted to a window containing the main tsunami signal, subsampled, and represented as a smooth function using B-splines. An OPE was then fitted to these functional outputs to obtain the predictive mean function and predictive uncertainty for each input $\boldsymbol{\theta}$. The OPE hyperparameters were tuned using held-out runs, and emulator checks, calibration details, sensitivity diagnostics, and simulation settings are deferred to the supplementary material.

The aggregate uncertainty scale in the denominator of the implausibility was calibrated in a data-driven manner for each observation site. Details of the calibration procedure and the resulting values are given in the supplementary material.
\subsection{What the functional criterion captures at the DART stations}
\label{subsec:result_dart}
%
%
%
%
%
\begin{figure}[b!]
\begin{center}
\includegraphics[width=6in]{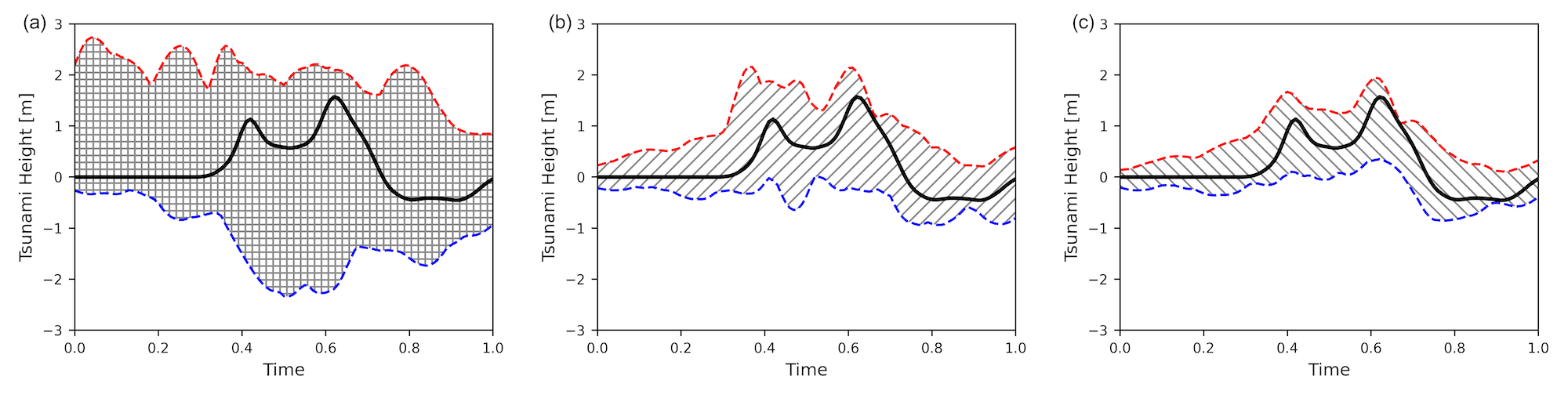}
\end{center}
\caption{\small Emulation and FHM results at DART 23228 in Wave 1. For each ensemble, the upper and lower envelopes are shown with the pseudo-observed series; the horizontal axis is rescaled to [0,1]. Panel (a): all $10^6$ candidate curves. Panel (b): curves retained under FHM using the original tsunami time series only (49,996 curves). Panel (c): curves retained under FHM using both the original tsunami time series and first-derivative information (18,292 curves).
\label{fig:pic_observed_DART}}
\end{figure}
At DART~23228, the station closest to the source region, we compare Wave~1 results based on the original tsunami time series alone with those obtained by combining the original time series with first-derivative information. 
When FHM was based only on the original tsunami time series, time series whose amplitudes and overall shapes were close to the pseudo-observed tsunami time series remained in the station-specific non-implausible set, but some curves with noticeable mismatches in the arrival times of peaks and troughs also remained. 
This indicates that the functional criterion based on the original tsunami time series primarily reflects amplitude differences and broad differences in overall shape, while being to some extent tolerant to temporal misalignment. 
By contrast, when derivative information was also incorporated, differences in the timing of growth and decay, as well as in rates of change, were more strongly emphasized, so that curves exhibiting phase misalignment were further excluded.
Thus, combining function values with derivative information improves discrimination not only with respect to amplitude but also with respect to timing. 
The same comparison was carried out for the remaining DART stations; the corresponding results are reported in the supplementary material.
\subsection{Shrinkage of the NROY region under multi-wave FHM}
\label{subsec:result_nroy}
%
%
%
%
%
\begin{figure}[t!]
\begin{center}
\includegraphics[width=6in]{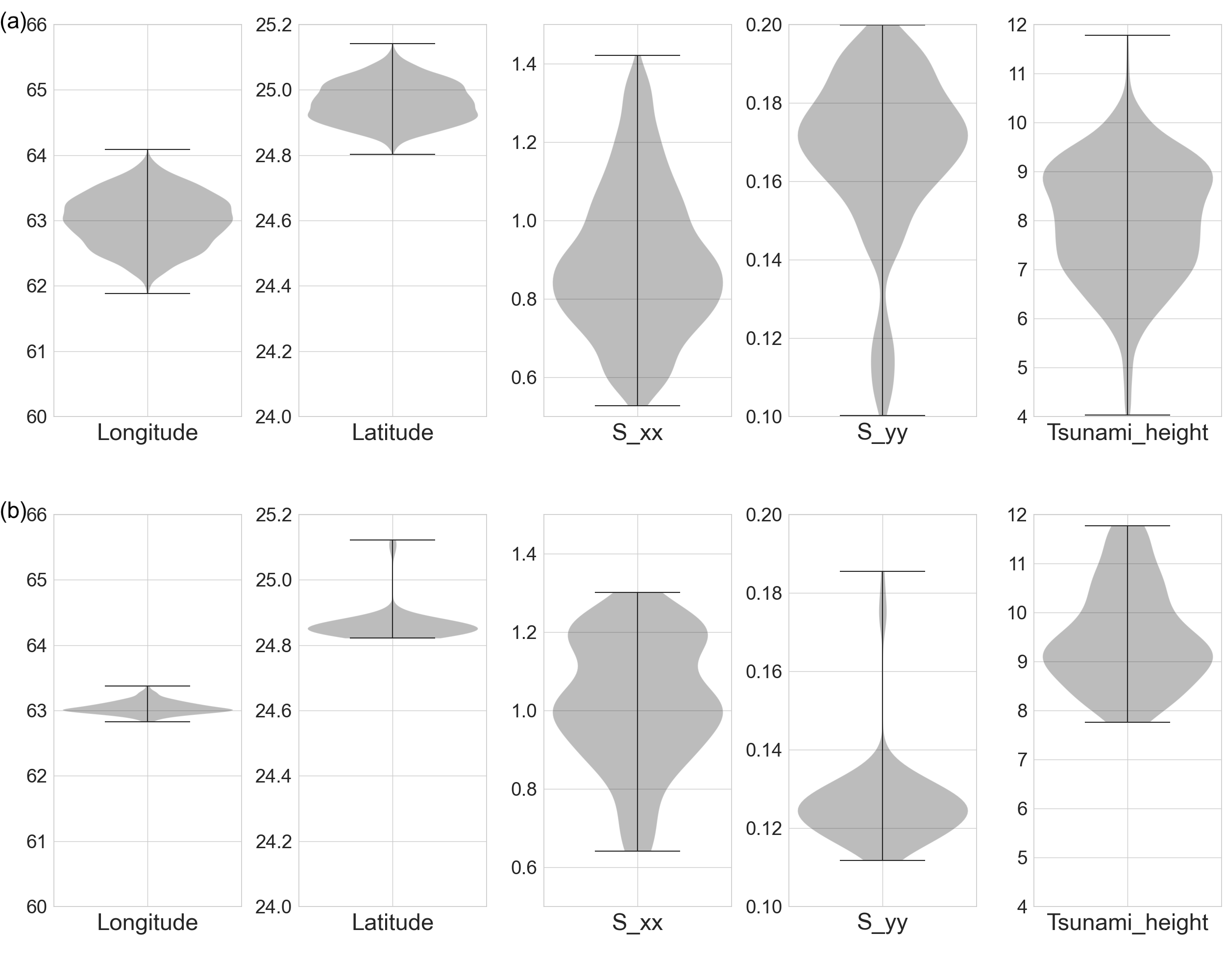}
\end{center}
\caption{\small Marginal distributions of the parameters in the NROY region, summarized by violin plots. Panel (a): Wave~1. Panel (b): Wave~2.
\label{fig:violin_plot}}
\end{figure}
At each HM wave, the final NROY set was defined as the intersection of the candidate inputs classified as non-implausible for both the
original and first-derivative series at all four DART stations. 
Figure~\ref{fig:violin_plot} summarizes the resulting contraction through the marginal distributions of each parameter in the NROY regions obtained after Wave~1 and Wave~2.

The results show that, already at Wave~1, the location parameters were substantially narrowed. 
Because the initial ranges for the location, shape, and amplitude parameters were deliberately chosen to be relatively broad, this finding indicates that FHM can effectively eliminate input regions that are inconsistent with the observations even under weak prior information, while leaving appreciable uncertainty in the dimensions governing source amplitude and shape.

At Wave~2, the design was updated over the reduced region identified in Wave~1 and FHM was repeated, yielding further contraction, particularly along the amplitude- and shape-related dimensions. 
In this case study, Wave~1 therefore primarily constrained the location parameters, whereas Wave~2 further tightened the remaining uncertainty in amplitude and shape. 
Implausibility values over parameter planes are shown in the supplementary material.
\subsection{Downstream prediction along the Mumbai coast}
\label{subsec:result_mumbai}
%
%
%
%
%
\begin{figure}[b!]
\begin{center}
\includegraphics[width=6in]{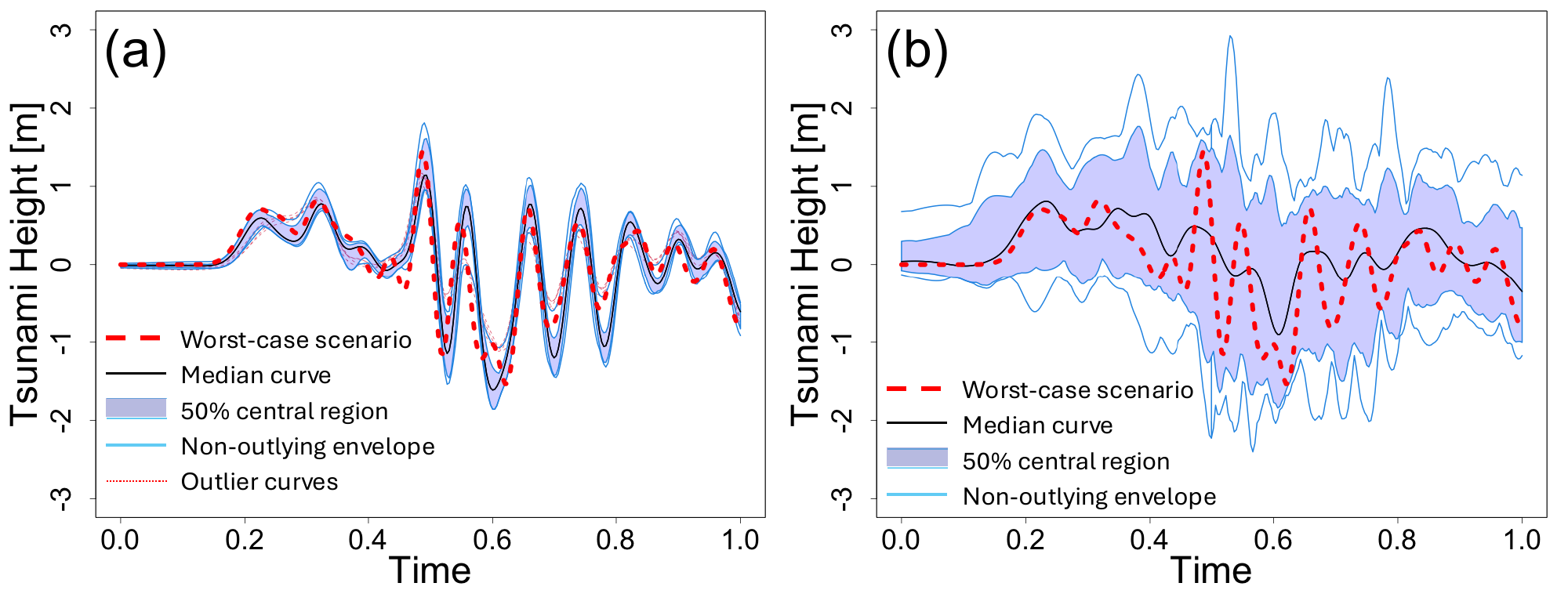}
\end{center}
\caption{\small Functional boxplot summaries of retained tsunami time-series predictions at Mumbai~001. Panel (a): FHM. Panel (b): landmark-based HM. In each panel, the solid black curve denotes the median, the shaded region denotes the $50\%$ central region, and the outer envelope gives the non-outlying range. The reference tsunami time series generated under the worst-case scenario is overlaid for comparison.
\label{fig:Mumbai_001}}
\end{figure}
Parameter samples from the Wave~2 NROY region were used to generate tsunami time-series predictions at four locations along the Mumbai coast. 
Since the NROY region was identified from the DART stations, the Mumbai sites provide an assessment of predictive performance at downstream locations that were not used in the screening. For visualization, the resulting predictive ensembles were summarized using functional boxplots \citep{Sun_2012}, which report the median curve, the $50\%$ central region, and the overall non-outlying envelope.

Figure~\ref{fig:Mumbai_001} shows the results for Mumbai~001, which we use as the representative coastal site in the main text. Under FHM, the median curve and central region closely follow the reference tsunami time series generated under the worst-case scenario, while the corresponding envelope remains relatively narrow over the main portion of the signal. This indicates that the reduced input region identified from the DART stations yields a concentrated set of coastal predictions while preserving the dominant temporal structure of the reference series.

For comparison, we also applied HM using four scalar landmark features from each tsunami time series: the maximum, the minimum, the time to the maximum, and the time to the minimum. We refer to this comparator as landmark-based HM, and implemented it using the same two-wave design. The Wave~2 FHM NROY set contained 132 candidate inputs, whose
corresponding tsunami curves were evaluated at Mumbai~001,
whereas landmark-based HM retained 38,631 candidate inputs.

Landmark-based HM does not adequately capture structure distributed across the full time series, including local attenuation, subsequent re-amplification, and later-stage shape differences. As a result, curves that agree with the selected landmarks but still deviate from the reference tsunami time series generated under the worst-case scenario over the full time domain remain in the retained set.

Taken together, the Mumbai~001 results show that, when the simulator output is intrinsically functional, FHM yields a sharper predictive summary than landmark-based HM. Similar results for Mumbai~002--004 are reported in the supplementary material.

%
%
%
%
%
\section{Conclusion and Discussion}
\label{sec:discussion}

In this paper, we proposed FHM, extending conventional HM from scalar or finite-dimensional vector outputs to functional outputs such as time series and spatial fields.
The central idea of FHM is to evaluate the discrepancy between an observed function and the emulator mean prediction as a functional measure of mismatch, and to normalize it by a scale that reflects observation error, model discrepancy, and emulator uncertainty. 
In this way, the basic principle of HM, the progressive elimination of input regions that are inconsistent with the observations, can be extended to functional outputs. 
The resulting framework provides a practical methodology for reducing the input space of expensive simulators while retaining information from the full functions.

From a methodological perspective, three aspects are particularly important. 
First, we made explicit the formulation of implausibility for functional outputs and introduced a Wiener-process-based random-projection distance-like score as the numerator of the mismatch criterion. 
Second, we adopted a multi-output Gaussian process perspective for functional emulation and implemented it through the OPE, thereby preserving dependence across the output domain while providing both predictive means and the predictive uncertainty required in the denominator of the implausibility. 
Third, rather than relying on strong assumptions about the shape of the implausibility distribution, we adopted a conservative, distribution-free threshold design based on Chebyshev's inequality, which stabilizes the implementation of HM for functional outputs. 
The resulting methodology integrates functional emulation, uncertainty quantification, and input-space screening within a single framework.

The case study based on the Makran tsunami scenario demonstrated the practical effectiveness of this framework. 
At the DART stations, incorporating first-derivative information in addition to the original tsunami time series led to the further exclusion of curves exhibiting mismatches in arrival timing, thereby illustrating the discriminatory advantage of using the full functional output. 
Moreover, under multi-wave FHM, the NROY region contracted progressively even under weak prior information, with uncertainty first reduced mainly in the location parameters and subsequently in amplitude- and shape-related dimensions. 
In downstream prediction along the Mumbai coast, the predictive summaries obtained from FHM reproduced the main structure of the reference tsunami time series generated under the worst-case scenario well and yielded more concentrated predictive bands than the landmark-based scalar-output HM benchmark. 
These findings indicate that, for simulator outputs with complex temporal structure, treating the output as a function rather than reducing it to a small number of summary features can provide both stronger discrimination and more informative prediction.

Several challenges nevertheless remain. 
In this paper, the aggregate uncertainty scale in the denominator of the implausibility was calibrated in a data-driven manner for each observation site, but a more systematic treatment of this choice remains desirable. 
A further limitation concerns the multi-wave design strategy. In practical HM, outputs entering the implausibility may be adapted across waves, with harder-to-emulate outputs introduced only after sufficient contraction of the NROY region \citep{vernon2018bayesian}. Here, for transparency, we fixed the functional-output construction across waves and generated the second-wave design from a rectangular approximation to the Wave 1 NROY region. Since NROY regions may be irregular or disconnected, more principled NROY-aware sampling strategies, including specialized sampling algorithms, evolutionary-Monte-Carlo designs, and subset-simulation-based approaches, remain important directions for future work \citep{andrianakis2017efficient, Williamson_2013arXiv, Gong_2021}.
Likewise, although the OPE was adopted here for computational efficiency, other multi-output or more flexible functional emulators may be preferable when the output dependence structure is more complex.
Accordingly, FHM should be viewed not as a method tied to a specific application domain, but as a general methodology for efficiently excluding input regions that are inconsistent with observations and for reducing expensive functional-output simulators to a scale at which downstream analysis becomes feasible. 
An important next step is therefore to apply FHM to other functional-output simulators, including those arising in climate, ocean, fluid, and structural-response applications, in order to develop clearer practical guidance on emulator choice, uncertainty calibration, distance design, and the number of HM waves.

\section{Disclosure statement}\label{disclosure-statement}

The authors have no conflict of interest to declare.

\section{Data Availability Statement}\label{data-availability-statement}
The code and data supporting this study will be made available in a GitHub
repository upon publication.







%

  \bibliography{bibliography.bib}

@article{biau2008performance,
  title={On the performance of clustering in Hilbert spaces},
  author={Biau, G{\'e}rard and Devroye, Luc and Lugosi, G{\'a}bor},
  journal={IEEE Transactions on Information Theory},
  volume={54},
  number={2},
  pages={781--790},
  year={2008},
  publisher={IEEE}
}

@article{cuesta2006random,
  title={Random projections and goodness-of-fit tests in infinite-dimensional spaces},
  author={Cuesta-Albertos, Juan Antonio and Fraiman, Ricardo and Ransford, Thomas},
  journal={Bulletin of the Brazilian Mathematical Society},
  volume={37},
  number={4},
  pages={477--501},
  year={2006},
  publisher={Springer}
}

@article{nieto2016topologically,
  title={A topologically valid definition of depth for functional data},
  author={Nieto-Reyes, Alicia and Battey, Heather},
  year={2016}
}

@article{cuesta2008random,
  title={The random Tukey depth},
  author={Cuesta-Albertos, Juan Antonio and Nieto-Reyes, Alicia},
  journal={Computational Statistics \& Data Analysis},
  volume={52},
  number={11},
  pages={4979--4988},
  year={2008},
  publisher={Elsevier}
}

@book{vempala2005random,
  title={The random projection method},
  author={Vempala, Santosh S},
  volume={65},
  year={2005},
  publisher={American Mathematical Soc.}
}

@article{cuesta2007sharp,
  title={A sharp form of the Cram{\'e}r--Wold theorem},
  author={Cuesta-Albertos, Juan Antonio and Fraiman, Ricardo and Ransford, Thomas},
  journal={Journal of Theoretical Probability},
  volume={20},
  number={2},
  pages={201--209},
  year={2007},
  publisher={Springer}
}

@article{andrianakis2017efficient,
  title={Efficient history matching of a high dimensional individual-based {HIV} transmission model},
  author={Andrianakis, Ioannis and McCreesh, Nicky and Vernon, Ian and McKinley, Trevelyan J and Oakley, Jeremy E and Nsubuga, Rebecca N and Goldstein, Michael and White, Richard G},
  journal={SIAM/ASA Journal on Uncertainty Quantification},
  volume={5},
  number={1},
  pages={694--719},
  year={2017},
  publisher={SIAM},
  doi = {https://doi.org/10.1137/16M1093008}
}

@article{vernon2014galaxy,
  title={Galaxy formation: Bayesian history matching for the observable universe},
  author={Vernon, Ian and Goldstein, Michael and Bower, Richard},
  journal={Statistical science},
  pages={81--90},
  year={2014},
  publisher={JSTOR}
}

@article{vernon2018bayesian,
  title={Bayesian uncertainty analysis for complex systems biology models: emulation, global parameter searches and evaluation of gene functions},
  author={Vernon, Ian and Liu, Junli and Goldstein, Michael and Rowe, James and Topping, Jen and Lindsey, Keith},
  journal={BMC systems biology},
  volume={12},
  number={1},
  pages={1--29},
  year={2018},
  publisher={BioMed Central}
}

@incollection{Craig1996,
    author = {Craig, P S and Goldstein, M and Seheult, A H and Smith, J A},
    isbn = {9780198523567},
    title = "{Bayes Linear Strategies for Matching Hydrocarbon Reservoir History}",
    booktitle = "{Bayesian Statistics 5: Proceedings of the Fifth Valencia International Meeting}",
    publisher = {Oxford University Press},
    year = {1996},
    month = {05},
    doi = {10.1093/oso/9780198523567.003.0004}}

@article{rougier2008efficient,
  title={Efficient emulators for multivariate deterministic functions},
  author={Rougier, Jonathan},
  journal={Journal of Computational and Graphical Statistics},
  volume={17},
  number={4},
  pages={827--843},
  year={2008},
  publisher={Taylor \& Francis}
}

@article{lay2005great,
  title={The great Sumatra-Andaman earthquake of 26 december 2004},
  author={Lay, Thorne and Kanamori, Hiroo and Ammon, Charles J and Nettles, Meredith and Ward, Steven N and Aster, Richard C and Beck, Susan L and Bilek, Susan L and Brudzinski, Michael R and Butler, Rhett and others},
  journal={science},
  volume={308},
  number={5725},
  pages={1127--1133},
  year={2005},
  publisher={American Association for the Advancement of Science}
}

@article{guillas2018functional,
author = {Serge Guillas and Andria Sarri and Simon J. Day and Xiaoyu Liu and Frederic Dias},
title = {{Functional emulation of high resolution tsunami modelling over Cascadia}},
volume = {12},
journal = {The Annals of Applied Statistics},
number = {4},
publisher = {Institute of Mathematical Statistics},
pages = {2023 -- 2053},
year = {2018},
doi = {10.1214/18-AOAS1142}}

@article{sarri2012statistical,
  title={Statistical emulation of a tsunami model for sensitivity analysis and uncertainty quantification},
  author={Sarri, Andria and Guillas, Serge and Dias, Frederic},
  journal={Natural Hazards and Earth System Sciences},
  volume={12},
  number={6},
  pages={2003--2018},
  year={2012},
  publisher={Copernicus Publications G{\"o}ttingen, Germany}
}

@article{marron2015functional,
  title={Functional data analysis of amplitude and phase variation},
  author={Marron, James Stephen and Ramsay, James O and Sangalli, Laura M and Srivastava, Anuj},
  journal={Statistical Science},
  pages={468--484},
  year={2015},
  publisher={JSTOR}
}

@article{pukelsheim1994three,
  title={The three sigma rule},
  author={Pukelsheim, Friedrich},
  journal={The American Statistician},
  volume={48},
  number={2},
  pages={88--91},
  year={1994},
  publisher={Taylor \& Francis}
}

@inproceedings{gonzalez1998deep,
  title={Deep-ocean assessment and reporting of tsunamis (DART): Brief overview and status report},
  author={Gonzalez, Frank I and Milburn, Hank M and Bernard, Eddie N and Newman, Jean C},
  booktitle={proceedings of the international workshop on tsunami disaster mitigation},
  volume={19},
  pages={2},
  year={1998},
  organization={NOAA Tokyo, Japan}
}

@article{william1984extensions,
  title={Extensions of Lipschitz mapping into Hilbert space},
  author = {Johnson, William B. and Lindenstrauss, Joram},
  journal={Contemporary mathematics},
  volume={26},
  pages={189-206},
  year={1984}
}

@article{dasgupta2013experiments,
  title={Experiments with random projection},
  author={Dasgupta, Sanjoy},
  journal={arXiv preprint arXiv:1301.3849},
  year={2013}
}

@article{cuesta2007random,
  title={The random projection method in goodness of fit for functional data},
  author={Cuesta-Albertos, Juan Antonio and del Barrio, Eustasio and Fraiman, Ricardo and Matr{\'a}n, Carlos},
  journal={Computational Statistics \& Data Analysis},
  volume={51},
  number={10},
  pages={4814--4831},
  year={2007},
  publisher={Elsevier}
}

@article{gopinathan2021probabilistic,
  title={Probabilistic quantification of tsunami current hazard using statistical emulation},
  author={Gopinathan, Devaraj and Heidarzadeh, Mohammad and Guillas, Serge},
  journal={Proceedings of the Royal Society A},
  volume={477},
  number={2250},
  pages={20210180},
  year={2021},
  publisher={The Royal Society Publishing}
}

@book{ramsay2005fitting,
  title={Fitting differential equations to functional data: Principal differential analysis},
  author={Ramsay, James O and Silverman, Bernard W},
  year={2005},
  publisher={Springer}
}

@article{baba2015parallel,
  title={{Parallel implementation of dispersive tsunami wave modeling with a nesting algorithm for the 2011 Tohoku tsunami}},
  author={Baba, Toshitaka and Takahashi, Narumi and Kaneda, Yoshiyuki and Ando, Kazuto and Matsuoka, Daisuke and Kato, Toshihiro},
  journal={Pure and Applied Geophysics},
  volume={172},
  pages={3455--3472},
  year={2015},
  publisher={Springer}
}

@article{percival2014automated,
  title={Automated tsunami source modeling using the sweeping window positive elastic net},
  author={Percival, Daniel M and Percival, Donald B and Denbo, Donald W and Gica, Edison and Huang, Paul Y and Mofjeld, Harold O and Spillane, Michael C},
  journal={Journal of the American Statistical Association},
  volume={109},
  number={506},
  pages={491--499},
  year={2014},
  publisher={Taylor \& Francis}
}

@article{wang2016functional,
  title={Functional data analysis},
  author={Wang, Jane-Ling and Chiou, Jeng-Min and M{\"u}ller, Hans-Georg},
  journal={Annual Review of Statistics and its application},
  volume={3},
  pages={257--295},
  year={2016},
  publisher={Annual Reviews}
}

@article{wang2021review,
  title={Review on recent progress in near-field tsunami forecasting using offshore tsunami measurements: {S}ource inversion and data assimilation},
  author={Wang, Y and Tsushima, H and Satake, K and Navarrete, P},
  journal={Pure and Applied Geophysics},
  volume={178},
  pages={5109--5128},
  year={2021},
  publisher={Springer}
}

@article{Sun_2012,
author = {Ying Sun and Marc G. Genton},
title = {Functional {B}oxplots},
journal = {Journal of Computational and Graphical Statistics},
volume = {20},
number = {2},
pages = {316--334},
year = {2011},
publisher = {Taylor \& Francis},
doi = {10.1198/jcgs.2011.09224}
}

@article{Polet_and_Kanamori_2000,
    author = {Polet, J. and Kanamori, H.},
    title = {Shallow subduction zone earthquakes and their tsunamigenic potential},
    journal = {Geophysical Journal International},
    volume = {142},
    number = {3},
    pages = {684-702},
    year = {2000},
    month = {09},
    issn = {0956-540X},
    doi = {10.1046/j.1365-246x.2000.00205.x},
    url = {https://doi.org/10.1046/j.1365-246x.2000.00205.x},
    eprint = {https://academic.oup.com/gji/article-pdf/142/3/684/6018301/142-3-684.pdf},
}

@ARTICLE{Okal1988,
  title    = "Seismic parameters controlling far-field tsunami amplitudes: {A} review",
  author   = "Okal, Emile A",
  journal  = "Natural Hazards",
  volume   =  1,
  number   =  1,
  pages    = "67--96",
  month    =  mar,
  year     =  1988
}

@article{LIU2018102,
title = {Remarks on multi-output Gaussian process regression},
journal = {Knowledge-Based Systems},
volume = {144},
pages = {102-121},
year = {2018},
issn = {0950-7051},
doi = {https://doi.org/10.1016/j.knosys.2017.12.034},
author = {Haitao Liu and Jianfei Cai and Yew-Soon Ong}
}

@ARTICLE{Williamson_2013arXiv,
       author = {{Williamson}, Daniel and {Vernon}, Ian},
        title = "{Efficient uniform designs for multi-wave computer experiments}",
      journal = {arXiv e-prints},
         year = 2013,
        month = sep,
          eid = {arXiv:1309.3520},
        pages = {arXiv:1309.3520},
          doi = {10.48550/arXiv.1309.3520},
archivePrefix = {arXiv},
       eprint = {1309.3520},
 primaryClass = {stat.ME},
       adsurl = {https://ui.adsabs.harvard.edu/abs/2013arXiv1309.3520W}
}

@article{Gong_2021,
	author  = {Z. T.  Gong and F. A.  DiazDelaO and Peter O. Hristov and Michael Beer},
	title   = {History Matching with Subset Simulation},
	journal = {International Journal for Uncertainty Quantification},
	issn    = {2152-5080},
	year    = {2021},
	volume  = {11},
	number  = {5},
	pages   = {19--38},
	DOI     = {10.1615/Int.J.UncertaintyQuantification.2021033543},
}

@article{seidman2025vppe,
  title={VPPE: Application of Scaled Vecchia Approximations to Parallel Partial Emulation},
  author={Seidman, Josh and Spiller, Elaine T},
  journal={arXiv preprint arXiv:2508.19144},
  year={2025}
}

@article{katzfuss2022scaled,
  title={Scaled Vecchia approximation for fast computer-model emulation},
  author={Katzfuss, Matthias and Guinness, Joseph and Lawrence, Earl},
  journal={SIAM/ASA Journal on Uncertainty Quantification},
  volume={10},
  number={2},
  pages={537--554},
  year={2022},
  publisher={SIAM}
}

@article{gu2016parallel,
  title={PARALLEL PARTIAL GAUSSIAN PROCESS EMULATION FOR COMPUTER MODELS WITH MASSIVE OUTPUT},
  author={Gu, Mengyang and Berger, James O},
  journal={The Annals of Applied Statistics},
  pages={1317--1347},
  year={2016},
  publisher={JSTOR}
}

@article{harbitz2014submarine,
  title={Submarine landslide tsunamis: how extreme and how likely?},
  author={Harbitz, Carl B and L{\o}vholt, Finn and Bungum, Hilmar},
  journal={Natural Hazards},
  volume={72},
  number={3},
  pages={1341--1374},
  year={2014},
  publisher={Springer}
}

\ifarXiv
  \includepdf[pages=-]{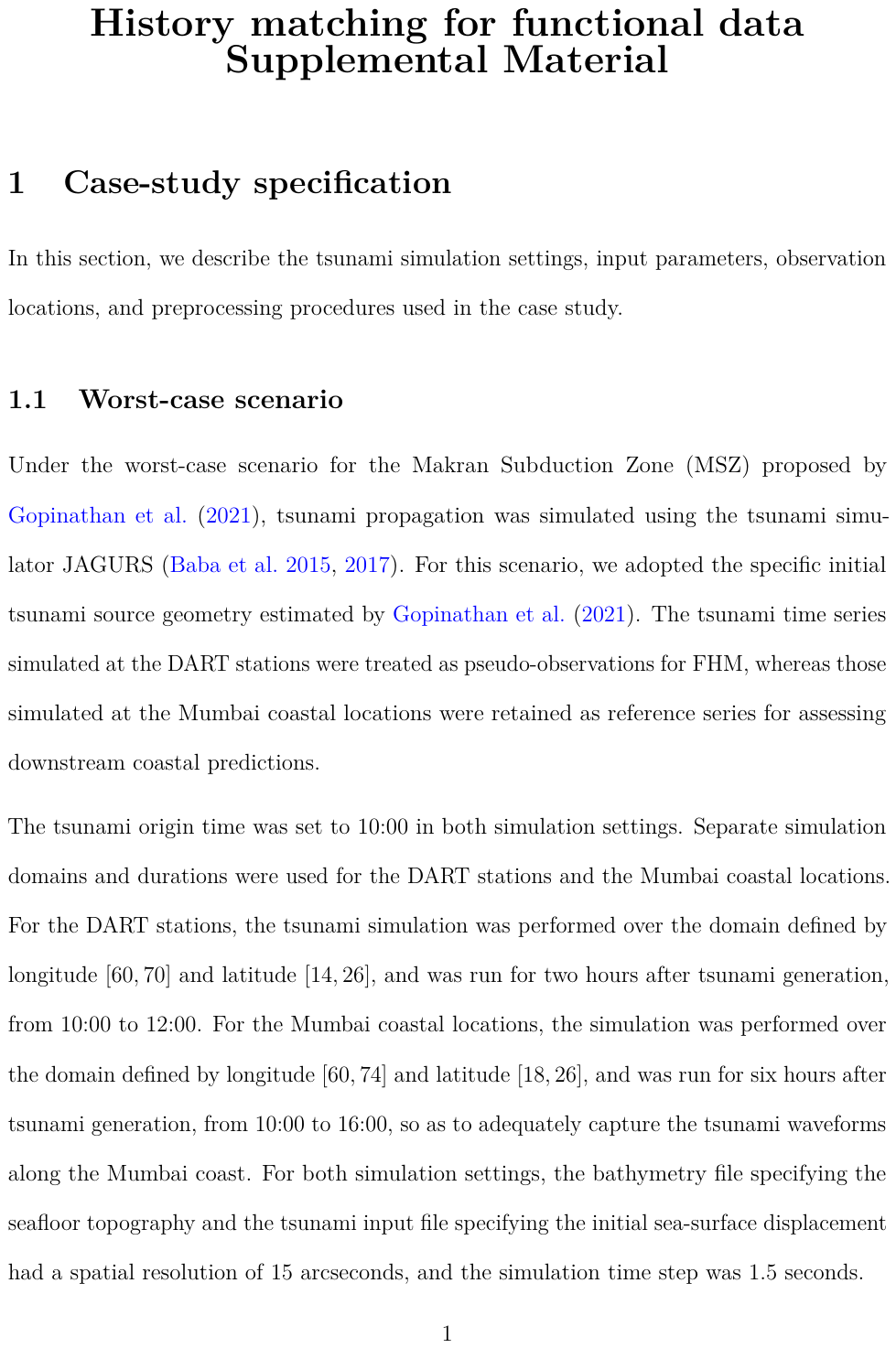}
\fi

\end{document}